\documentclass[a4paper,11pt]{article}

\usepackage{jcappub} 

\usepackage[T1]{fontenc} 

\title{Initial Conditions of Inhomogeneous Universe
and the Cosmological Constant Problem}

\author{Tomonori Totani}
\affiliation{Department of Astronomy, 
The University of Tokyo, Hongo, Tokyo 113-0033, Japan}

\emailAdd{totani@astron.s.u-tokyo.ac.jp}

\abstract{Deriving the Einstein field equations (EFE) with matter
  fluid from the action principle is not straightforward, because mass
  conservation must be added as an additional constraint to make
  rest-frame mass density variable in reaction to metric
  variation. This can be avoided by introducing a constraint
  $\delta(\sqrt{-g}) = 0$ to metric variations $\delta g^{\mu\nu}$,
  and then the cosmological constant $\Lambda$ emerges as an
  integration constant. This is a removal of one of the four
  constraints on initial conditions forced by EFE at the birth of
  the universe, and it may imply that EFE are unnecessarily
  restrictive about initial conditions. I then adopt a principle that
  the theory of gravity should be able to solve time evolution
  starting from arbitrary inhomogeneous initial conditions about
  spacetime and matter. The equations of gravitational fields
  satisfying this principle are obtained, by setting four auxiliary
  constraints on $\delta g^{\mu\nu}$ to extract six degrees of freedom
  for gravity. The cost of achieving this is a loss of general
  covariance, but these equations constitute a consistent theory if
  they hold in the special coordinate systems that can be uniquely
  specified with respect to the initial space-like hypersurface when
  the universe was born.  This theory predicts that gravity is
  described by EFE with non-zero $\Lambda$ in a homogeneous patch of
  the universe created by inflation, but $\Lambda$ changes
  continuously across different patches. Then both the smallness and
  coincidence problems of the cosmological constant are solved by the
  anthropic argument.  This is just a result of inhomogeneous initial
  conditions, not requiring any change of the fundamental physical
  laws in different patches. }

\begin{document}
\maketitle
\flushbottom

\section{Introduction}
\label{sec:intro}

The energy density of vacuum appears as the cosmological constant
term, $\Lambda$, in the Einstein field equations (EFE).  Non-zero
$\Lambda$ in the present universe was already implied by some
observations in early 1990's
\cite{Efstathiou+90,Fukugita+90,Yoshii_93,Krauss+95,Ostriker+95},
which were further strengthened by type Ia supernova data
\cite{Riess+98,Perlmutter+99}, and finally confirmed by the WMAP data
of the cosmic microwave background (CMB) \cite{Spergel+03}.  The
energy density inferred from the observed value $\Lambda_{\rm obs}$ is
smaller than the Plank energy density $(c^7/\hbar G^2)$ by a factor of
$10^{120}$, and smaller than those expected by particle physics
theories by at least $10^{60}$, which is the smallness problem of the
cosmological constant.  It is even more complicated by the coincidence
problem that we are living in a very special epoch when the matter
density becomes comparable with that of vacuum.  In spite of a large
number of proposals, there is no compelling candidate of the solution
(see \cite{Frieman+08,Caldwell+09,Martin_12,Weinberg+13} for recent
reviews on theories and observations).

In this work a new theory of gravity, which includes general
relativity as a particular case, is proposed to solve the cosmological
constant problem, based on reconsideration of the metric degrees of
freedom (DOFs) in the action principle to derive the equations of
gravitational fields, which are related to constraints on initial
conditions.  In \S \ref{section:const_volume}, we start by an
examination of the EFE derivation with fluid matter, pointing out a
rather strange aspect that the fluid rest-mass density must be varied
at the same time with metric $g_{\mu\nu}$ to obtain the matter energy
tensor.  This can be avoided by introducing a condition of constant
covariant volume element, i.e., $\delta(\sqrt{-g}) = 0$, in variation
of the metric tensor, $\delta g^{\mu\nu}$.  Then $\Lambda$ appears in
EFE as an integration constant.  This is often called the unimodular
theory\footnote{The unimodular condition is normally defined as
  $\sqrt{-g} = 1$, but in this work we only require the variational
  condition of $\delta (\sqrt{-g}) = 0$. It is always possible to make
  a general coordinate transformation to a system in which $\sqrt{-g}
  = 1$ everywhere \cite{vanderBij+82}.}  or equivalently the
trace-free EFE, and has been discussed in the literature for a long
time, starting from Einstein himself
\cite{Einstein_19a,Einstein_19b,Einstein_19c,Anderson+71,vanderBij+82,Buchmuller+88,Unruh_88,Henneaux+89,Ng+90,Ng+91,Finkelstein+00,Ng+01,Smolin_09,Ellis+11,Ellis_13}.
The motivations to introduce this constraint depend on authors: a
theory for the internal structure of the electron 
\cite{Einstein_19a,Einstein_19b,Einstein_19c},
a fundamental atomic length \cite{Anderson+71}, a viewpoint of the
little group in the description of massless spin-two particles
\cite{vanderBij+82}, quantum gravity \cite{Unruh_88}, or the
cosmological constant problem
\cite{Einstein_19a,Einstein_19b,Einstein_19c,Henneaux+89,Ng+90,Ng+91,Ng+01,Smolin_09,Ellis+11}.
However, to the author's knowledge, the motivation described in this
work is not found in previous studies.

The constant $\sqrt{-g}$ condition does not solve the cosmological
constant problem, but gives an important hint. The true physical DOFs
for gravity are at most six in the ten components of $g_{\mu\nu}$,
after removing four DOFs of coordinate transformation.
Correspondingly, four of the ten components of EFE are not dynamical
equations, but constraints to initial conditions. This is a result of
varying the action with all the ten metric components, while the
physical DOFs of gravity is less than ten. This may indicate that EFE
are unnecessarily restrictive about initial conditions.  The constant
$\sqrt{-g}$ condition indeed removes one of these constraints.  In
\S\ref{section:PFIC}, we extend this, and introduce a principle that
the gravity theory must be able to solve time evolution for arbitrary
initial conditions of inhomogeneous spacetime and matter.  In
conventional general relativity, only a special universe satisfying
the four constraints in EFE is allowed to start. It may be fascinating
that a theory of spacetime can predict time evolution starting from
any physically possible initial states about spacetime and matter at
the birth of the universe. The equations for gravitational fields
satisfying this principle will be presented in \S
\ref{section:new_eq}, by introducing three more constraints on $\delta
g^{\mu\nu}$. The cost of getting this is violation of general
covariance. However, these equations still give a consistent theory to
determine spacetime evolution, if we consider that these equations
hold in a uniquely specified coordinate system defined by the initial
space-like hypersurface at the birth of the universe.  Though the
proposed theory loses the beauty of general covariance, it acquires
another property that no constraints need to be forced to the initial
conditions. Furthermore, as discussed in \S \ref{section:implication},
the cosmological constant problems are solved simply as a result of
inhomogeneous initial conditions and subsequent inflation.  General
covariance is restored and EFE with a non-zero $\Lambda$ apply in a
homogeneous patch of the universe created by inflation.

The sign convention in this work is the same as that of
\cite{bakoten}, and the fundamental constants $c$ and $G$ will be set
equal to unity.  Greek indices run from 0 to 3, while Latin indices
are for spatial indices running from 1 to 3.  Partial derivatives are
represented by colons or $\partial_\mu$, while covariant derivatives
by semicolons or $\nabla_\mu$.

\section{Examination of the Action Principle in
Relativistic Cosmology}
\label{section:const_volume}

EFE can be derived from the principle of least action with the simple
Lagrangian density for gravity [$L_G = -R/(2 \kappa)$, where $R$ is
  the Ricci scalar and $\kappa = 8\pi$].  To derive EFE with perfect
fluid, the energy-momentum tensor, $T_{\mu\nu} = (\rho + p) u_\mu
u_\nu - p g_{\mu\nu}$, must be derived by metric variation of the
matter Lagrangian $L_M$, where $\rho$ and $p$ are energy density and
pressure in the fluid rest-frame, respectively, and $u^\mu$ is fluid
four-velocity.  It is possible to find such $L_M$, but not quite
simple.  Here, curl-free fluid of non-relativistic matter ($p=0$ and
$\rho = \rho_m$) is considered for simplicity, where $\rho_m$ is the
rest mass density in the fluid rest-frame, though it is possible to
extend to the case of non-zero pressure and vorticity
\cite{Schutz_70,Brown_93,Poplawski_08,Fukagawa+10}.  The fundamental
matter Lagrangian is simply given by $L_{M, f} = - \rho_m$, but
additional conditions of the identity $g_{\mu\nu}u^\mu u^\nu = 1$ and
conservation of rest mass [$\nabla_\mu(\rho_m u^\mu) = 0$] are
necessary with Lagrange multipliers of $\zeta$ and $\eta$, resulting
in the following action:
\begin{eqnarray}
S &=& \int \left( L_G + L_M \right) \sqrt{-g} \, d^4x \ , \\
L_M &=& L_{M, f} + \zeta \, (g_{\mu\nu}u^\mu u^\nu - 1) 
  + \eta \, \nabla_\mu( \rho_m u^\mu ) \ .
\label{eq:action}
\end{eqnarray}
Variations about $\rho_m$, $u^\mu$, $\zeta$, and $\eta$, but fixing
$g_{\mu\nu}$, result in the mass conservation and the Euler equation
of motion, $u^\mu \nabla_\mu u^\nu = 0$, under appropriate boundary
conditions.  The Lagrange multipliers are also determined by this
process, as $\zeta = - \rho_m / 2$ and $\partial_\mu \eta = - u_\mu$.

Now consider variation of the action $S$ about metric $g^{\mu\nu}$.
To get EFE, the energy momentum tensor must be derived by the standard
formula (e.g., \cite{bakoten}):
\begin{eqnarray}
T_{\mu\nu} = \frac{2}{\sqrt{-g}} \left[
\frac{\partial (L_M \sqrt{-g}) }{
\partial g^{\mu\nu}} - \frac{\partial}{\partial x^\lambda}
\frac{\partial (L_M \sqrt{-g}) }{\partial g^{\mu\nu}_{\ ,\lambda}}
\right] \ ,
\end{eqnarray}
and indeed we find $T_{\mu\nu} = \rho_m u_\mu u_\nu$, but only if the
same Lagrange multipliers ($\zeta$ and $\eta$) determined by the
variations about fluid quantities are used.  The mass conservation
constraint includes the Christoffel symbols $\Gamma^\mu_{\rho\sigma}$,
and the term coming from variation of $\delta \Gamma^\mu_{\mu\lambda}
= - \partial_\lambda (g_{\mu\nu} \ \delta g^{\mu\nu}/2)$ has a role to
cancel the unnecessary term of $\propto \rho_m \, g_{\mu\nu}$ generated
by variation of the determinant, $\delta \sqrt{-g} = - \sqrt{-g} \,
g_{\mu\nu} \, \delta g^{\mu\nu} /2$.  However, it seems rather unbalanced
that the Lagrange multipliers are determined solely by fluid dynamics
and these must be used also in variation about $g^{\mu\nu}$. If metric
is determined by the action principle for fixed fields of $\rho_m$ and
$u^\mu$, the Lagrange multipliers need not be the same as those
determined by variation about fluid quantities.  In this case the
constraints in Lagrangian reduce DOFs of the metric variations $\delta
g^{\mu\nu}$, and correspondingly increased DOFs of solutions appear as
the freedom of choosing the Lagrange multipliers. Instead, if we
require the same multipliers for variations of $g^{\mu\nu}$ and fluid
quantities, this is equivalent to requesting that the action is
stationary when both the metric and fluid quantities are varied
simultaneously satisfying the constraints.  In this case all the ten
metric components can be varied independently without reduction of
DOFs, because the constraints can be met by accordingly changing fluid
quantities.  This is reasonable for $\zeta$, because variations of
$g^{\mu\nu}$ necessarily change the proper time $ds$ and hence
four-velocity $u^\mu = dx^\mu/ds$ for a world line $x^\mu(s)$ of a
fluid element.

However, the use of the same $\eta$ means that $\rho_m$ may also
change accordingly with $\delta g^{\mu\nu}$ to satisfy the mass
conservation condition.  This can be seen more clearly in the
Friedmann-Lema\^itre-Robertson-Walker (FLRW) metric in cosmology,
where the $ii$ component of EFE is derived from variation about
$g^{ii} = -a^{-2}$ keeping $g^{00} = 1$, where $a$ is the scale
factor.  The mass conservation can be written as $\nabla_\mu( \rho_m
u^\mu ) = \dot \rho_m + \Gamma^\mu_{\mu 0} \, \rho_m = 0$, where
$\Gamma^\mu_{\mu 0} = 3 \dot a / a$ and the dot denotes time
derivative. Then, if $\rho_m(t)$ is fixed, the Hubble parameter $H
\equiv \dot a / a$ cannot be varied, but in the flat universe case the
only physically meaningful quantity about spacetime is $H(t)$. This
means that independent variations of $g^{00}$ and $g^{ii}$, which are
necessary to derive the full set of EFE in FLRW cosmology, are
possible only when $\rho_m$ is allowed to vary with $a$ to meet the
mass conservation, i.e., $\delta(\rho_m a^3) = 0$. Alternatively, if
we request that $\rho_m$ is fixed, there must be a constraint to the
relation between $\delta g^{00}$ and $\delta g^{ii}$.

The source of gravity in EFE to determine the spacetime curvature is
energy density rather than a total mass in a volume element, and it
seems strange if $\rho_m$ also needs to change with metric
variation. The equivalence principle tells us that gravity force can
be erased in a local inertial frame by an appropriate coordinate
transformation, and hence the gravity should not affect the physical
quantities observed in a local inertial frame including
$\rho_m$. Therefore I here adopt a principle that rest-frame density
of conserving mass (or particle number) should not be altered in the
process of finding solutions of gravitational fields by the action
principle.  This means that variations of $\delta g^{\mu\nu}$ lose one
DOF to meet the mass conservation, and we should leave the multiplier
$\eta$ as a free field in the solution of gravitational fields.  If we
split as $\eta \equiv \eta_f + \eta_g$, where $\eta_f$ is that
determined by fluid variation (i.e., $\partial_\mu \eta_f = - u_\mu$),
the term including $\eta_f$ is absorbed into $T_{\mu\nu}$ and the
right-hand-side of EFE is modified as
\begin{eqnarray}
R_{\mu\nu} - \frac{1}{2} R \, g_{\mu\nu} &=& \kappa \, T_{\mu\nu} 
 + \Lambda(x^\mu) g_{\mu\nu} \ , \\
\Lambda(x^\mu) &\equiv& \kappa \, \rho_m u^\lambda \, 
  \partial_\lambda \eta_g \ ,
\end{eqnarray}
where $R_{\mu\nu}$ is the Ricci tensor.
Because of the contracted Bianchi identity
and the energy conservation of fluid ($\nabla_\mu T^{\mu\nu} = 0$), we
find $\partial_\mu \Lambda(x^\mu) = 0$, i.e., $\Lambda$ must be a
constant everywhere in the spacetime.  Therefore the change from the
original EFE is just adding a cosmological constant, but this is now
an integration constant depending on boundary or initial conditions,
rather than a fundamental physical constant.

However, the above scheme is still not satisfactory as a general
theory to determine the spacetime geometry for fixed matter
quantities, because the constraint on metric variation is introduced
as the form of mass conservation.  The mass conservation is not always
satisfied in reality (e.g., fluid composed of decaying particles into
radiation). Another constraint depending only on the metric itself but
resulting in the same equation would be better, and in fact we can
find one.  Both the metric variations of $\sqrt{-g}$ and
$\nabla_\mu(\rho_m u^\mu)$ have a form of $\propto g_{\mu\nu} \delta
g^{\mu\nu}$, and hence if we introduce the constraint
$\delta(\sqrt{-g}) = 0$ instead of mass conservation, the same result
is obtained.  The condition of fixed $\sqrt{-g}$ is reasonable if we
request that $\rho_m$ is fixed against variation of $\delta
g^{\mu\nu}$, because $\rho_m \sqrt{-g} \, d^4x$ is also a scalar
quantity and the only way to keep these two scalars constant is to fix
$\sqrt{-g}$.  I reached the idea of introducing this constraint purely
by the motivation described above. Then I learned, in surveying the
literature, that this constraint has been discussed for a long time
(see references in Introduction), though motivations were different.
In the following it is assumed that this constraint is essential when
solutions of gravitational fields are found by the action principle.

\section{The Principle of Free Initial Condition}
\label{section:PFIC}

The outcome of introducing the constant $\sqrt{-g}$ condition is
interesting because $\Lambda$ emerges as an integration constant, but
it is not satisfactory for the cosmological constant problem, given
that there is no guiding principle to determine its value as an
initial condition.  However, $\Lambda$ as an integration constant has
a different nature from $\Lambda$ being a fundamental physical
constant, because it increases one DOF for allowed solutions, and a
wider set of initial conditions become possible.  In FLRW cosmology
with $\Lambda$ as a fundamental physical constant, the Friedmann
equation sets a constraint on the allowed initial condition; the
Hubble parameter is simply related to energy density at any time in a
flat universe.  However, if $\Lambda$ is added as an arbitrary
integration constant, it can easily be erased by taking a linear
combination of the $00$ and $ii$ components of EFE, resulting in a
second-order time differential equation for the scale factor $a$,
allowing any combination of $H = \dot a / a$ and $\rho$ as an initial
condition.  However, this is only for the FLRW metric, and we cannot
start with an arbitrary initial condition in an inhomogeneous
universe, because $\Lambda$ is a universal constant throughout the
spacetime.

In general relativity, the constraints on initial conditions come from
the ten components of EFE derived from the action principle with
variations of all the ten metric components $\delta g^{\mu\nu}$
independently, though the physical DOFs of gravity are at most six
because of the four DOFs of general coordinate transformations.  In
EFE, the second order time derivatives of the $0\mu$ metric
components, $\partial_0^2 g_{0\mu}$, do not appear, and the $0\mu$
components of the Einstein tensor $G_{\mu\nu} = R_{\mu\nu} -
R \, g_{\mu\nu} / 2$ do not include the second-order time derivative of
any metric component. Therefore it is natural to regard the spatial
components $g_{ij}$ as the dynamical variables, and we can always
choose a synchronous coordinate system in which $g_{00} = 1$ and
$g_{0i} = 0$ everywhere for any spacetime geometry.  Then the six $ij$
components of EFE determine the evolution of $g_{ij}$, and the
additional four $0\mu$ components of EFE are not dynamical equations
but just give constraints on initial conditions.  (If they are met at
the initial time, the contracted Bianchi identity ensures that they
hold at any time.)  Therefore time evolution of spacetime cannot be
solved for arbitrary $g_{ij}$, $\partial_0 g_{ij}$, and $T_{\mu\nu}$
at an initial space-like hypersurface.

Considering physical DOFs of gravitational fields, we may not need to variate
the action with all the ten metric components independently.  In fact,
the constant $\sqrt{-g}$ condition removes one DOF of $\delta
g^{\mu\nu}$, and as a result one of the four constraint equations
disappears.  One may still consider that variation should be done with
all the ten metric components, since the action should be stationary
not only against $\delta g^{\mu\nu}$ by six DOFs of gravity, but also
against those induced by coordinate transformation. However, metric
variations generated by infinitesimal coordinate transformation,
$x^\mu \rightarrow x^\mu + \xi^\mu$, are $\delta g^{\mu\nu} =
\xi^{\mu; \nu} + \xi^{\nu; \mu}$.  If we consider $\xi^\mu$ that is
nonzero in an infinitesimal region but zero otherwise, the stationary
action condition results in the contracted Bianchi identity and energy
conservation after integrating by parts, as written in textbooks
(e.g., \cite{bakoten}).  These are constraints on the derivatives of
EFE rather than EFE themselves.  Hence there is no particular reason
to require the stationary action against all the ten metric
components.

Therefore the standard 10-component EFE may be unnecessarily
restrictive as the theory of gravity, and a theory with less constraints
on initial conditions may be constructed. We can physically imagine an
arbitrary matter distribution embedded in an arbitrary spacetime
structure at the initial space-like hypersurface when the universe was
born.  Though such a universe is not allowed in the standard EFE, it
would be fascinating if we have a theory that can predict time
evolution for such a universe as well.  Here I propose the principle
of free initial condition: {\it the gravity theory must be formulated
  so that time evolution can be solved for arbitrary initial conditions
  about spacetime and matter.}  The initial conditions of spacetime
should be those for the metric and its time derivative, like many
dynamical systems described in physics.  The conventional EFE
obviously do not satisfy this principle.  Introducing the constant
$\sqrt{-g}$ condition is not yet satisfactory, because the principle
is satisfied only in isotropic and homogeneous
FLRW cosmology.  Clearly, we need to add
three more constraints on $\delta g^{\mu\nu}$ to remove the remaining
three constraint equations.

\section{Gravitational Field Equations for Arbitrary Initial Conditions}
\label{section:new_eq}

It is reasonable to expect that the three more constraints on $\delta
g^{\mu\nu}$ are related with the $0i$ components of EFE, and hence the
synchronous coordinate system in which $g_{00} = g^{00} = 1$ and
$g_{0i} = g^{i0} =0$ seems a natural reference frame where the gravity
DOFs are easy to treat. Consider time evolution from a space-like
hypersurface defined by $x^0 = t_i$.  The direction of the normal at
each point on this hypersurface is uniquely determined, and $g_{0i} =
g^{0i} = 0$ if we set the spatial coordinate so that $x^i$ do not
change along this direction. Therefore setting $g_{0i} = g^{0i} = 0$
is just a matter of coordinate choice, which is unrelated to the
physical DOFs of gravity.  If we set the constraints of $\delta g^{0i}
= 0$ on the metric variations, the $0i$ components of EFE do not need
to hold.  However, we cannot set $\delta g^{00} = 0$ because it
includes a gravity DOF, though it is not independent from $\delta
g^{ij}$ because of the constant $\sqrt{-g}$ condition. The six DOFs of
gravity are then represented by seven-component variations of $\delta
g^{00}$ and $\delta g^{ij}$ with the constraint of $\delta(\sqrt{-g})
\propto g_{00} \delta g^{00} + g_{ij} \delta g^{ij} = 0$.  Requiring
that the action is stationary against these variations, we find
\begin{eqnarray}
R_{\mu\nu} - \frac{1}{2} R \, g_{\mu\nu} = \kappa \, T_{\mu\nu} 
+ \Lambda_0 \, g_{\mu\nu} + \Xi_{\mu\nu} \ , 
\label{eq:totani}
\end{eqnarray}
where
\begin{eqnarray}
\Xi_{\mu\nu} = \left( 
\begin{array}{c|ccc} 
0 & \Lambda_1 & \Lambda_2 & \Lambda_3 \\
\hline
\Lambda_1 &&& \\
\Lambda_2 & \multicolumn{3}{c}{ 
\raisebox{-1ex}[0pt][0pt]{\LARGE 0} } \\
\Lambda_3 &&& \\
\end{array}
\right) \ .
\label{eq:Xi}
\end{eqnarray}
Here, $\Lambda_0(x^\mu)$ is a Lagrange multiplier field corresponding
to the constant $\sqrt{-g}$ condition, and $\Lambda_i(x^\mu)$ are
arbitrary fields to make the $0i$ components of EFE ineffective.  Of
course, the set of $(\Lambda_0, \Lambda_i)$ is not a vector, but
$\Lambda_0$ is a scalar and $\Lambda_i$ are components\footnote{The
  character $\Xi$ has been chosen because $\Xi$ has three non-zero
  components and it looks similar to the Chinese character ``three''.}
of $\Xi_{\mu\nu}$.  The contravariant version $\Xi^{\mu\nu}$ also
satisfies $\Xi^{00} = \Xi^{ij} = 0$, and their nonzero components are
denoted as $\Xi^{0i} = \Xi^{i0} = \Lambda^i = g^{00} g^{ij}
\Lambda_j$.

However, a problem of this equation as a theory for gravitational
fields is that it violates general covariance, because the above
condition $\delta g^{0i} = 0$ is dependent on the choice of coordinate
systems, in contrast to the constant $\sqrt{-g}$ condition.
Nevertheless, general covariance is not necessarily indispensable for
a theory to determine spacetime evolution. If we can specify a unique
coordinate system for a given spacetime, equations that hold only in
such a system can be a consistent theory. Here we take this option.
However, the synchronous condition is not sufficient to specify such a
system, because it does not uniquely specify the coordinate system,
the Lorentz transformation being an example. The form of
$\Xi_{\mu\nu}$ does not keep eq. (\ref{eq:Xi}) by a Lorentz
transformation, and hence eqs. (\ref{eq:totani}) with
eq. (\ref{eq:Xi}) cannot be a general form for any synchronous
coordinate systems.

To avoid this problem, it is assumed that any spacetime realized in
nature has a finite space-like physical boundary into the past as an
initial condition.  We do not know how a spacetime is born, but at
least the only one example that we know, i.e., our universe, seems to
satisfy this. If we can define a physically unique coordinate with
respect to this initial space-like hypersurface, and if equations hold
only within it, they become a consistent theory to predict time
evolution of spacetime starting from the initial hypersurface, even if
they violate general covariance.  We define a synchronized time $x^0 =
t_s$ on the initial hypersurface, and the spacetime does not extend to
the region of $x^0 < t_s$. If we set a spatial coordinate system at
$x^0 = t_s$, the synchronous coordinate system starting from this is
uniquely determined throughout this spacetime.  Only transformations
within spatial coordinates [$x'^i = f^i(x^j)]$ are allowed to keep the
synchronous condition at any point in the spacetime and $x^0 = t_s$ on
the initial hypersurface.  In such a set of coordinate systems, the
form of $\Xi_{\mu\nu}$ in eq. (\ref{eq:Xi}) is unchanged. [It is also
  kept against a transformation including only time coordinate, $x'^0
  = f^0(x^0)$]. Therefore, if eqs. (\ref{eq:totani}) and (\ref{eq:Xi})
are assumed to hold only in these coordinate systems, they give a
consistent theory to predict time evolution.

Then the principle of free initial conditions is now satisfied.  The
evolution of gravitational fields $g_{ij}$ is determined by
eq. (\ref{eq:totani}) with the form of $\Xi_{\mu\nu}$ given in eq.
(\ref{eq:Xi}).  The seven equations of the $00$ and $ij$ components
include $\Lambda_0(x^\mu)$, and this can be erased resulting in six
second-order time differential equations for $g_{ij}$.  Then we can
take any combination of $g_{ij}$, their time derivatives, and
$T_{\mu\nu}$ as an initial condition.  The initial values of
$\Lambda_0$ and $\Lambda_i$ are also determined by the initial
conditions of spacetime and matter.  In EFE, energy conservation of
matter is automatically satisfied by the contracted Bianchi identity,
but here the least action condition of matter must be independently
required to ensure $\nabla_\mu T^{\mu\nu} = 0$. Then four-divergence
of eq. (\ref{eq:totani}) gives four first-order time differential
equations to determine evolution of $\Lambda_0$ and $\Lambda^i$ as
\begin{eqnarray}
\nabla_\mu ( \Lambda_0 g^{\mu\nu} + \Xi^{\mu\nu} ) = 0 \ .
\label{eq:Lambda_ev}
\end{eqnarray}
Obviously this includes the standard general relativity ($\Lambda_0 =
\Lambda_i = 0$), and hence this is an extension of general relativity.

Though this theory violates general covariance, eqs. (\ref{eq:totani})
is written in a generally covariant form, and hence these equations
can be extended to any coordinate systems in the same form, if we
define the ordinary tensor transformation law for $\Xi_{\mu\nu}$.  It
violates general covariance in the sense that the expression of
$\Xi_{\mu\nu}$ takes a special form of eq. (\ref{eq:Xi}) in a special
set of coordinate systems.  Violation of general covariance for a
theory of gravity may not be unreasonable if we consider the following
points. It is standard to start with the Lagrangian density of gravity
being the Ricci scalar, $L_G = - R/(2\kappa)$, but a peculiar aspect
is that $R$ includes the second-order time derivatives of
$g_{\mu\nu}$, while normally Lagrangians include up to first-order
derivatives of dynamical variables. It is possible to derive EFE
starting from a Lagrangian that does not include second-order time
derivatives \cite{bakoten}, by defining
\begin{eqnarray}
Q \equiv g^{\rho \sigma}(\Gamma^\mu_{\rho \nu}\Gamma^\nu_{\sigma \mu} 
  - \Gamma^\nu_{\rho \sigma}\Gamma^\mu_{\nu \mu})
\end{eqnarray}
which is related to $R$ as
\begin{eqnarray}
R \sqrt{-g} = Q \sqrt{-g} + 
\frac{\partial}{\partial x^\mu} \left[W^\mu \sqrt{-g} \right] \ ,
\end{eqnarray}
where
\begin{eqnarray}
W^\mu \equiv g^{\rho \sigma} \, \Gamma^\mu_{\rho \sigma} 
  - g^{\mu \rho} \, \Gamma^\sigma_{\rho \sigma} \ .
\end{eqnarray}
Therefore the action principle gives the same result if we take $L_G =
- Q/(2\kappa)$ and variations of $W^\mu$ are fixed to zero at
integration boundaries.  Another good point of this Lagrangian is the
quantities that should be fixed at boundaries.  In the case of the
conventional Lagrangian, variations of $\Gamma^\mu_{\rho\sigma}$ that
include time derivatives of $g_{\mu\nu}$ must be fixed to zero at
boundaries, because
\begin{eqnarray}
\delta[ R \sqrt{-g} \, ] = \delta( g^{\mu \nu} R_{\mu \nu} \sqrt{-g} )
= G_{\mu\nu} \, \delta g^{\mu \nu} \sqrt{-g} + \frac{\partial}{\partial
  x^\mu} (w^\mu \sqrt{-g}) \ ,
\end{eqnarray}
where
\begin{eqnarray}
w^\mu = g^{\rho \sigma} \, \delta \Gamma^\mu_{\rho \sigma} 
  - g^{\mu \rho} \, \delta \Gamma^\sigma_{\rho \sigma} \ .
\end{eqnarray}
However, in ordinary steps to derive the Euler-Lagrange equation of
motion from the action principle, first-order time derivatives of
dynamical quantities (in this case $\Gamma^\mu_{\rho\sigma}$) need not
be fixed at boundaries.  If we take $L_G \propto Q$, indeed we need to
fix only $\delta g^{\mu\nu}$ to zero at boundaries, because
\begin{eqnarray}
\delta[Q \sqrt{-g} \, ] &=& G_{\mu\nu} \, \delta g^{\mu\nu} \sqrt{-g} 
- \, \frac{\partial}{\partial x^\mu} \left[ 
(\delta g^{\rho \sigma} \, \Gamma^\mu_{\rho \sigma} 
  - \delta g^{\mu \rho} \, \Gamma^\sigma_{\rho \sigma}) 
\sqrt{-g} + W^\mu \, \delta (\sqrt{-g})
\right] \ 
\end{eqnarray}
and the four-divergence term includes only $\delta g^{\mu\nu}$ without
$\delta \Gamma^\mu_{\rho\sigma}$. Then $Q$ seems a more natural
Lagrangian density than $R$ to derive the Einstein tensor
$G_{\mu\nu}$, but $Q$ is not a scalar, and the action becomes
dependent on the global properties of a specified coordinate
system. Lagrangian density should be closely related to the energy
density, and $Q$ has a similar form (quadratic in
$\Gamma^\mu_{\rho\sigma}$) to the energy-momentum psudotensor of
gravitational fields, which is not a tensor.  These considerations
imply that gravity is essentially global and dependent on coordinate
systems, and it may not be surprising that the constraints on $\delta
g^{\mu\nu}$ to extract six physical DOFs of gravity are expressed in a
simple form in special coordinate systems determined by the initial
space-like hypersurface.

\section{Implications for the Cosmological Constant Problem}
\label{section:implication}

The extended theory of gravity proposed here allows the cosmological
``constant'' $\Lambda_0$ to change on the initial space-like
hypersurface. In accordance with the standard paradigm of the
inflationary universe
\cite{Starobinsky_79,Kazanas_80,Guth_81,Sato_81,Linde_81,Albrecht+82},
here it is assumed that the universe started with a highly
inhomogeneous condition.  Then $\Lambda_0 g_{\mu\nu}$ and
$\Xi_{\mu\nu}$ would have fluctuations with amplitudes similar to that
of matter energy tensor $\kappa T_{\mu\nu}$ at that time.  To realize
an isotropic and homogeneous universe as observed today, inflation
must occur at least in some regions in the whole universe.  Let
$\Lambda_\phi = \kappa \rho_\phi$ be the cosmological constant
corresponding to the vacuum energy density $\rho_\phi$ of the inflaton
field $\phi$.  Quantitative conditions for successful inflation must
be investigated numerically, but we expect that there are some regions
where $\Lambda_\phi \, g_{\mu\nu}$ is dominant compared with
$\Lambda_0 g_{\mu\nu} + \Xi_{\mu\nu}$ in eq. (\ref{eq:totani}). Then
such regions would start inflation, and $\Lambda_0$ and $\Lambda^i$
evolve by eqs. (\ref{eq:Lambda_ev}).  Intuitively, $\Lambda^i$ should
become zero asymptotically, and consequently $\Lambda_0$ becomes a
constant by eqs. (\ref{eq:Lambda_ev}), because the universe becomes
isotropic and homogeneous by exponential expansion.

This can be examined more quantitatively if the background metric is
nearly isotropic. We can treat $\Lambda_0$ and $\Lambda^i$
perturbatively in the flat FLRW metric, and eqs. (\ref{eq:Lambda_ev}) become
\begin{eqnarray}
\partial_0 \Lambda_0 + \partial_i \Lambda^i &=& 0 
  \label {eq:Lambda1} \\
- \frac{\partial_i \Lambda_0}{a^2} + \partial_0 \Lambda^i
+ 5 H \Lambda^i &=& 0 \label{eq:Lambda2} \ .
\end{eqnarray}
From these two equations we can derive the equation for $\Lambda_0$ as
\begin{eqnarray}
\ddot \Lambda_0 + 5H \, \dot \Lambda_0 + \frac{\Delta \Lambda_0}{a^2}
 = 0 \ ,
\end{eqnarray}
where $\Delta$ is the Laplacian in the comoving coordinate.  The
characteristic inhomogeneity scale of $\Lambda_0$ would be the horizon
scale $H^{-1}$ just before inflation, and consider a Fourier mode of
this scale with a comoving wavenumber $|\vec k_{ic}| \sim H$, i.e.,
$\Delta \Lambda_0 = - k_{ic}^2 \Lambda_0$, where we define $a=1$ at
the beginning of inflation.  Once the inflation starts at $t=t_{si}$,
the evolution is $a = \exp[H(t-t_{si})]$ with a nearly constant $H$,
and this scale should soon become super-horizon.  Though $|\Lambda_0|$
is accelerated to a larger value by the force $k_{ic}^2
\Lambda_0/a^2$, the friction force $-5H \dot \Lambda_0$ limits the
rate of $\Lambda_0$ evolution as $\dot \Lambda_0 / \Lambda_0 \lesssim
(k_{ic}^2/5a^2H) \sim H a^{-2}$, and hence $\Lambda_0$ would become a
constant. Eq. (\ref{eq:Lambda2}) indicates that $\Lambda^i$ should
decay as $\propto \exp(-5Ht)$ when $ 5H |\Lambda^i| \gtrsim k_{ic}
|\Lambda_0|/a^2$, and hence $a |\Lambda^i|/|\Lambda_0|$ is limited to
be $\lesssim k_{ic}/(5aH) \sim a^{-1}$.  This means that
$\Xi^{\mu\nu}$ is negligible compared with $\Lambda_0 g^{\mu\nu}$
after inflation.  Then finally $\Lambda_0 g^{\mu\nu} + \Xi^{\mu\nu}$
would nearly become a cosmological constant term $\Lambda_{0,f} \,
g^{\mu\nu}$ in a homogeneous patch of the universe created by
inflation, where $\Lambda_{0,f}$ is a universal constant within the
patch, but its value can be positive or negative and changes
continuously on the comoving scale of $k_{ic}$. At the end of the
inflation, the contribution $\Lambda_\phi$ disappears, and the final
effective cosmological constant in a patch is $\Lambda_f \equiv
\Lambda_{0,f} + \Lambda_{\rm vac}$, where $\Lambda_{\rm vac} (= \kappa
\rho_{\rm vac} \ll \Lambda_\phi)$ is the microscopic vacuum energy
density in the universe today from any contributing sources including
quantum zero-point fluctuations.

The value of $\Lambda_{\rm vac}$ is expected to be much larger than
$\Lambda_{\rm obs}$ (i.e., the smallness problem), but the amplitude
of $\Lambda_{0,f}$ fluctuation would be even larger.  Then there
should be regions where $| \Lambda_f | \lesssim \Lambda_{\rm obs}$ in
a successfully inflated portion of the universe. This is the region
that is habitable for an intelligent life, because patches with
$\Lambda_f \ll - \Lambda_{\rm obs}$ should have collapsed much earlier
than the present epoch, and formation of structure and galaxies does
not occur in patches with $\Lambda_f \gg \Lambda_{\rm obs}$ because of
too fast expansion.  Therefore the smallness problem is solved by the
anthropic argument \cite{Weinberg_87}.  The comoving width of the
regions of $| \Lambda_f | \lesssim \Lambda_{\rm obs}$ is much smaller
than the initial inhomogeneity scale $k_{ic}^{-1}$ by a factor of
$\sim \Lambda_{\rm obs}/\Lambda_{0,f}$, and the total number of
$e$-foldings from the beginning to the end of inflation must be
sufficiently large to make this width much larger than the present-day
Hubble horizon.  The fractional change $\delta \Lambda_f / \Lambda_f$
is of order unity within the regions of $|\Lambda_f| \lesssim
\Lambda_{\rm obs}$, but those of $\Lambda_{0,f}$ and any other
physical quantities (e.g., matter density, properties of inflation,
and density fluctuation amplitude) are negligibly small.

It is expected that $\Lambda_f$ changes linearly with a spatial
position within the regions of $|\Lambda_f| \lesssim \Lambda_{\rm
  obs}$, because $\Lambda_{\rm obs}$ is much smaller than the typical
fluctuation amplitude of $\Lambda_{0,f}$. Then the prior probability
distribution function $P_{\rm pri}(\Lambda_f)$ should be almost
constant per unit $\Lambda_f$. The probability distribution to observe
$\Lambda_f$ should be $P_{\rm pri}$ multiplied by the efficiency of
creating intelligent life, $\epsilon_{\rm life}(\Lambda_f)$. It has
been shown that, under the assumption of constant $P_{\rm pri}$ and
estimating $\epsilon_{\rm life}(\Lambda_f)$ by galaxy formation
efficiency, the probability for us to observe $\Lambda_f \sim
\Lambda_{\rm obs}$ is not extremely small
\cite{Efstathiou_95,Martel+98}. Moreover, the probability of finding
$|\Lambda_f| \ll \Lambda_{\rm obs}$ is small because $P_{\rm pri}(<
|\Lambda_f|) \propto |\Lambda_f|$, and hence the coincidence problem
is also solved.

This picture is analogous to the concentration of human population to
coastal areas on Earth, if we regard $\Lambda_f$ as altitude above the
sea level; we cannot live under the sea level, while it is hard to
live in high positive altitude regions.  Therefore I call this the
coastal universe hypothesis.

\section{Discussion}

The coastal universe scenario predicts that the present-day
cosmological constant $\Lambda_f$ should vary at different positions,
on the comoving scale of initial inhomogeneity before inflation.
However, if inflation is sufficient, the expected change within the
Hubble horizon should be negligible, and therefore a more practical
prediction is that the observed universe should be described exactly
by the standard $\Lambda$CDM model.  The high precision experimental
tests on general relativity on the solar system scales are not
affected.  The current small $\Lambda_{\rm obs}$ is a result of huge
cancellation between $\Lambda_{0,f}$ and $\Lambda_{\rm vac}$, and the
latter should gravitate. This may be tested experimentally, e.g., by
examination of gravitational properties of the Casimir energy
\cite{Martin_12}.  Since the proposed theory changes DOFs of
gravitational fields, implications for quantum gravity theory would be
interesting. Gravitational wave background radiation would be
generated during inflation by quantum fluctuation of metric, and the
prediction by the proposed theory may be different from the standard
one because of the DOFs of $\Lambda_0 \, g_{\mu\nu} + \Xi_{\mu\nu}$,
though this becomes a cosmological constant by inflation at the
classical level. Examination of such prediction is beyond the scope of
this work, but this may be tested by the $B$-mode polarization of CMB
or direct detection experiments in future.

Though the coastal universe hypothesis uses the anthropic argument,
the change of $\Lambda_f$ across different homogeneous patches of the
universe is just a result of inhomogeneous initial conditions under
the same fundamental physical laws. Physical quantities other than
$\Lambda_f$ do not change, and $\Lambda_f$ should have a constant
prior probability distribution $P_{\rm pri}$ per unit $\Lambda_f$.
This is in contrast to some other explanations of $\Lambda_{\rm obs}$
based on the anthropic argument (e.g., string landscape
\cite{Susskind_03}), in which not only $\Lambda$ but also other
physical quantities (like density fluctuation amplitude) and even the
fundamental physical laws or constants may also change.  The
anthropic argument may not simply work in these cases
\cite{Tegmark+98,Aguirre_01,Graesser+04}.  It is also highly uncertain
whether a constant $P_{\rm pri}(\Lambda_f)$ is realized in some other anthropic
scenarios \cite{Garriga+00}.  Astronomy is still showing a remarkable
development by large projects in wide wavelength ranges, and our
understanding of galaxy formation, star and planet formation, and even
the origin of life will be further improved in the future. Then we may
be able to calculate the observational probability distribution of
$\Lambda_f$ with a more realistic estimate of $\epsilon_{\rm
  life}(\Lambda_f)$, giving a more quantitative test of the coastal
universe scenario that varies only $\Lambda_f$ with a flat prior
probability distribution.

\end{document}